\documentclass[%
 reprint,
superscriptaddress,
 amsmath,amssymb,
 aps,
]{revtex4-2}

\usepackage{graphicx}
\usepackage{dcolumn}
\usepackage{bm}

\begin{document}

\preprint{APS/123-QED}

\title{Skyrmion elongation, duplication and rotation by spin-transfer torque under spatially varying spin current}

\author{Floris van Duijn}
\affiliation{Imec, B-3001 Leuven, Belgium}
\affiliation{KU Leuven, ESAT, B-3001 Leuven, Belgium}
\author{Javier Osca}
\affiliation{Imec, B-3001 Leuven, Belgium}
\affiliation{KU Leuven, ESAT, B-3001 Leuven, Belgium}
\author{Bart Sor\'{e}e}
\email[Author to whom correspondence should be addressed.   E-mail: ]{Bart.Soree@imec.be}
\affiliation{Imec, B-3001 Leuven, Belgium}
\affiliation{KU Leuven, ESAT, B-3001 Leuven, Belgium}
\affiliation{Universiteit Antwerpen, Departement Fysica, B-2000 Antwerpen, Belgium}


\date{\today}

\begin{abstract}
The effect of the spatially varying spin current on a skyrmion is numerically investigated. It is shown that an inhomogeneous current density induces an elongation of the skyrmion. This elongation can be controlled using current pulses of different strength and duration. Long current pulses lead to a splitting that forms two replicas of the initial skyrmion while for short pulses the elongated skyrmion relaxes back to its initial circular state through rotation in the MHz-GHz frequency range. The frequency is dependent on the strength of the damping coefficient.

\end{abstract}

\maketitle


\section{\label{sec:level1}Introduction}
Due to the limitations of Moore's law for CMOS, there is a need for novel devices that replace or assist current technologies. Skyrmion based devices are promising candidates for memory and logic applications due to their small size and stability compared with other spintronic proposals \citep{Logic, Tomasello}. Skyrmions are topologically protected magnetic structures, that were originally found in bulk crystals with a non-centrosymmetrical lattice \cite{muhlbauer} and their existence can be explained by the balance between exchange and Dzyaloshinskii-Moriya interaction (DMI) \citep{Dzyal, Moriya}. Later on, skyrmions were also observed in ultrathin magnetic films, where the inversion symmetry is broken at the interface between a ferromagnet and a heavy metal. Magnetic films are grown on heavy metals with a large spin-orbit coupling, which induces DMI at the interface \cite{bode} and can stabilize these skyrmions. 

In this research, we focus mainly on these ultrathin films, because an interesting property of skyrmions at interfaces is that they can easily be moved by spin polarized currents. This motion is driven by either spin-transfer torque (STT) \cite{STT1,STT2,STT3} or spin orbit torque (SOT) \cite{SOT, SOT3, SOT2}. 

We explore, using micromagnetic modelling, the conditions under which the skyrmion splits, stretches and rotates under the effect of STT caused by a spatially varying spin current in a 2-dimensional geometry, such as the interface of an ultrathin magnetic film with a heavy metal layer.

Duplication of skyrmions has been researched before in \cite{Logic, duplication1} with the help of simulations of confining geometries and tilted magnetic pulses in the range of 200 ps, that elongate the skyrmion as well. Here we show an alternative method for skyrmion duplication based on pulses of a spatially varying current. This current goes on opposite directions at each side of the skyrmion. After the applied current pulse, the skyrmion will elongate and eventually split into two versions of the original skyrmion, provided the pulse is long enough. On the other hand, for short current pulses, the skyrmion will relax back to its initial radius while rotating. These two phenomena may be the foundation of novel devices that can generate skyrmions or be used as oscillators.

This paper is organized as follows. In Sec. \ref{sec:level2} we start with a description of the model and an explanation of the employed method in this study. Section \ref{sec:level3} presents our results on the splitting of the skyrmion. We also discuss the stretching and rotation of the skyrmion in Sec. \ref{sec:level4}. The conclusion is given in Sec. \ref{sec:conclusion}.


\begin{figure}[h]
\includegraphics[width=1\columnwidth]{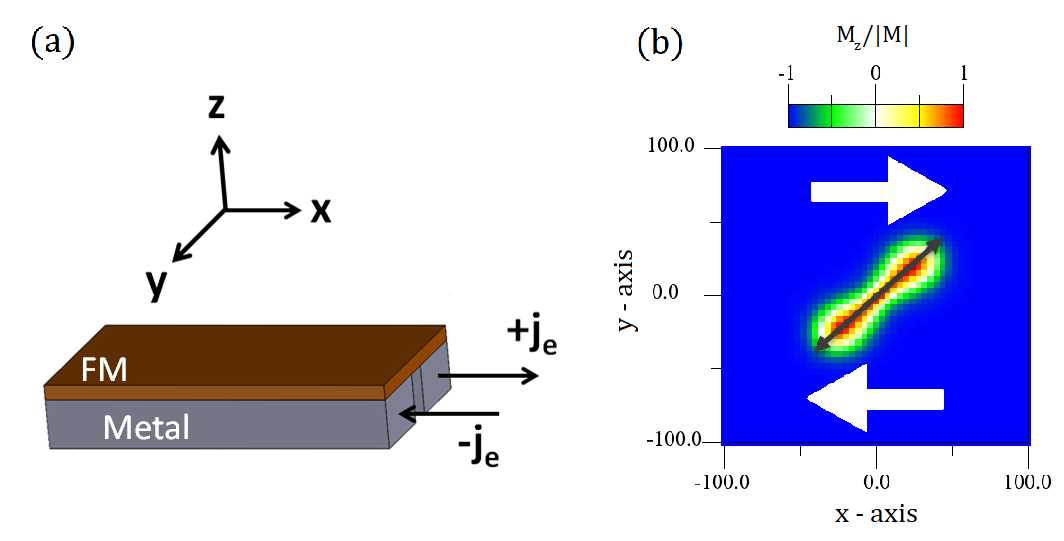}
\caption{(a) Device model schematic. We consider a device made of a ferromagnet build over two heavy metal layers separated by a gap where the spin current flows in opposite directions. (b) Elongated skyrmion through the effect of counter-propagating spin currents. Spin current directions for $y\geq 0 $ and $y<0$ are shown schematically with white arrows. The current pulse used, is $3\cdot10^{12}$ A/m$^2$ for 0.67 ns with device parameters $A= 2.5$ pJ/m, $D_{\bot}=1$ mJ/m$^2$, $\mu_0H = 0.1$ T $P=0.4$, $M_s=1000$ kA/m and $\alpha=0.015$. In order to determine the length of the skyrmion elongation in the wider axis (dark gray arrow in the figure), we consider the skyrmion boundaries to be located at $M_z/|M| = -0.8$.  }
\label{spin}
\end{figure}

\section{Model}\label{sec:level2}
We will consider a ferromagnetic-heavy metal interface where the metal layer is separated into two sub-layers with the spin currents flowing in opposite directions, as is shown in Fig. \ref{spin}(a). We model the ferromagnet-heavy metal interface as a 2D grid where the ferromagnet magnetization is defined on each grid point. The magnetization evolution is governed by the Landau-Lifshiftz-Gilbert equation (LLG):
\begin{equation}\label{eq:LLG}
\frac{d\textbf{m}}{dt} = -\textbf{m}\times(\gamma\mu_0\textbf{H}_{\rm{eff}} + \textbf{H}_{\rm{torq}}) +\alpha\textbf{m}\times\frac{d\textbf{m}}{dt}
\end{equation}
where \textbf{m} = \textbf{M}/M$_s$ is the normalized magnetization; M$_s$ is the saturation magnetization; $\gamma$ is the gyromagnetic ratio; $\mu_0$ is the magnetic permeability of free space; $\alpha$ is the damping constant and \textbf{H}$_{\rm{eff}}$ is the effective field. Note that the normalized magnetization is always one and each of the components vary from -1 to 1. The effective field is comprised of the exchange interaction, magnetic anisotropy, external magnetic field and DMI effective fields. Here, we focus only on the interfacial DMI \cite{dmi}, given by  
\begin{equation}
    \textbf{H$_{\rm{DMI}}$} = -(2D/\mu_0M_s)[(\nabla\times\textbf{m})_z -\nabla\textbf{m}_z]
\end{equation}
with $D$ the parameter taking into account the intensity of the DMI. The exchange interaction and anisotropy are given by the following two expressions
\begin{equation}
    \textbf{H}_{\textrm{exc}} = \frac{2}{\mu_0}\frac{J}{M_s}\nabla^2 \textbf{m}\textrm{ \  , \  } \textbf{H}_{\textrm{ani}} = \frac{2}{\mu_0}\frac{K}{M_s}\textbf{m}_z
\end{equation}

with $J$ and $K$ respectively the parameter intensity of the exchange effective field and anisotropy effective field terms. Furthermore, the torque term of the effective field from Eq. \ref{eq:LLG} is computed as:
\begin{equation}
\textbf{H$_{\rm{torq}}$} = \textbf{m}\times \left(\textbf{u}\cdot\nabla)\textbf{m}\right) + (\xi\textbf{u}\cdot\nabla)\textbf{m}
\end{equation}

where $\textbf{u} = j_e\mu_BP/eM_s(1+\xi^2) $. Here $j_e,\mu_B, P, e, M_s, \xi$ are respectively, the current density, the Bohr magneton, current polarization, electron charge. $\xi$ models the strength of the electronic diffusion in the metal. $\xi$ is zero if the current is transported ballistically and different from zero if electron scattering is present \citep{Torque}. 

In order to model the two heavy metal layers, a current profile is considered This current profile is achieved with the help of a shape function that varies the strength of the of the $x$-component of the current in the transverse direction. The $y$-component is considered to be zero. In addition, this current strength may be varied in time to create finite sized current pulses. 
\begin{equation}\label{shape}
\textbf{j}_x(y) = \textbf{j}_{x0} \left(2H(y)-1\right)\ ,
\end{equation}
with $\textbf{j}_x(y)$ the spatially varying spin current, $\textbf{j}_{x0}$ is a constant value equal to the maximum of the current density and $H(y)$ the Heaviside step function. The sign of $\textbf{j}_x$ indicates the direction of the current. This creates a model in which one section of the metal sub-layer of the simulation has a spin current flowing in the positive longitudinal direction of the device and in the other part flowing towards the negative one as shown in Fig. \ref{spin}(b). The skyrmion is located in a way that each half of its surface is impinged by a current of opposite sign. 

We simulate skyrmion dynamics in samples with the size of 300 x 300 nm$^2$ and a mesh size of 2 x 2 nm$^2$. The LLG equation is numerically integrated with the Euler method. In order for the ferromagnet-metal interface to hold a stable isolated skyrmion, the original parameters were obtained from \cite{Tomasello}. To allow for the skyrmion to elongate at lower current magnitudes, the values of $J$ and $D$ are reduced to one eighth of the original values.  The following parameters were chosen; exchange interaction $J = 2.5$ pJ/m, Dzyaloshinskii-Moriya interaction $D = 1$ mJ/m$^2$, polarization $P = 0.4$ and damping $\alpha = 0.015$. 

Theoretically, a skyrmion has an infinite radius, however, in our research the length is determined at the point where the normalized magnetization $M_z/M_s$ = -0.8 as illustrated in Fig. \ref{spin}(b) with the black arrow.

\section{Skyrmion deformation and splitting under current}\label{sec:level3}
\begin{figure}[h]
\includegraphics[width=0.8\columnwidth]{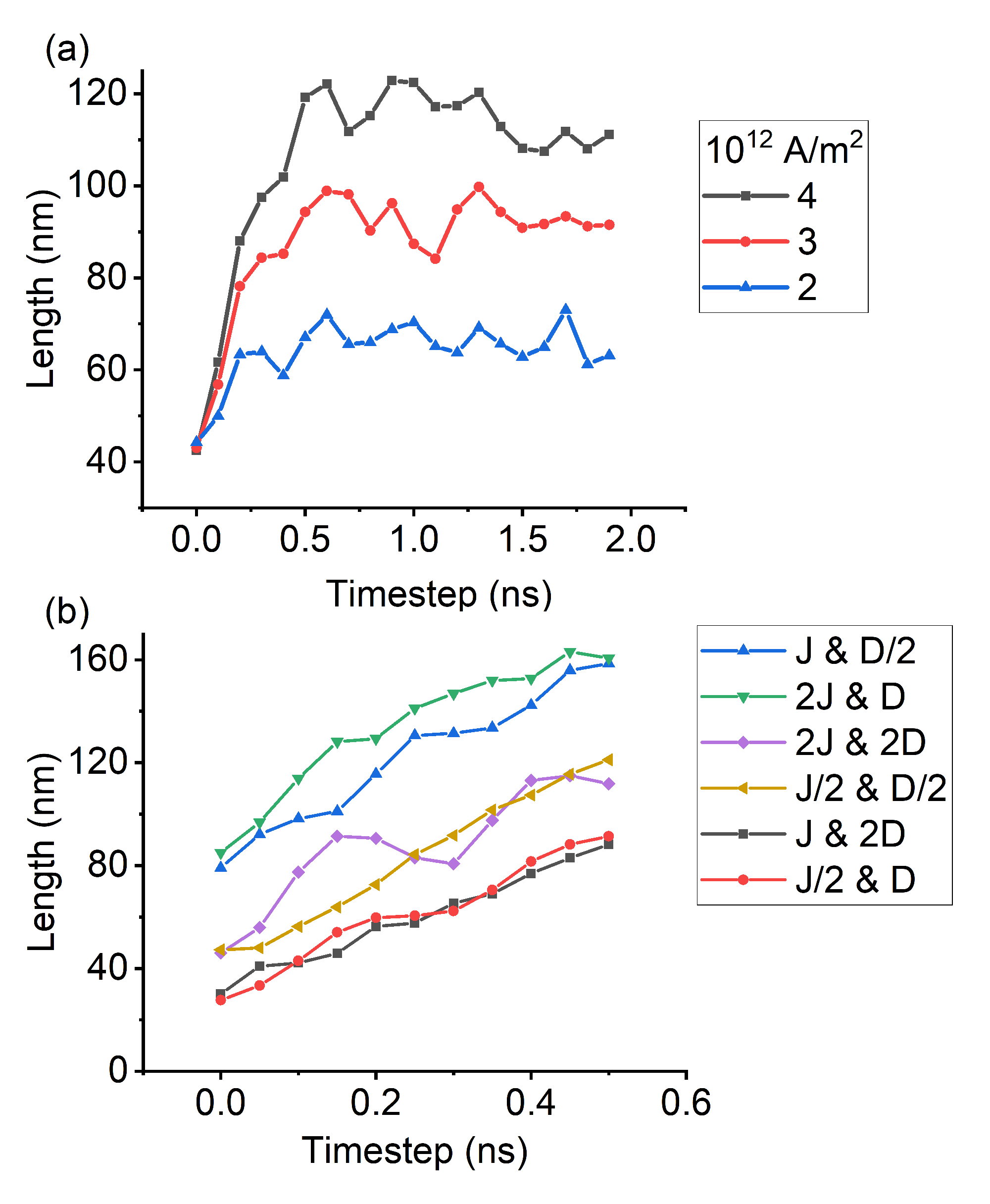}
\caption{(a) Skyrmion elongation evolution in time as a function of a single inhomogeneous current pulse of 0.5 ns. (b) Skyrmion elongation in time as a function of an infinite pulse for different LLG parameters, where $J$ and/or $D$ is halved or doubled. LLG parameters are $J = 2.5$ pJ/m, $D = 1$ mJ/m$^2$, $
\mu_0H = 0.1$ T  and $\alpha$ = 0.015.}
\label{2a}
\end{figure}

\subsection{Deformation with current pulses\label{lab:deform}}
Applying a spatially varying spin current on a centered skyrmion with respect to the two counter-propagating spin currents, such as in Figure \ref{spin}(b), generates two forces facing different directions. If this force is strong enough, the skyrmion begins to elongate indefinitely. This elongation is not stretching the skyrmion, but rather expanding it in size, hence the area is not conserved. The rate at which the skyrmion grows depends on the magnitude of the applied current. Therefore we can control the elongation of the skyrmion by switching off the current at a certain moment, creating finite current pulses. The final skyrmion elongation length after the current pulse, will also depend on the strength of the current pulse as shown in Fig. \ref{2a}(a). After the current pulse, the skyrmion is left to relax, if the elongation is not too large, the skyrmion starts to diminish in size over time until the skyrmion becomes circular again. Note too, that for small currents the skyrmion elongation shrink slightly as can be seen at $t=0.3$ ns. The skyrmion does not return to equilibrium smoothly, but wobbles as can be seen for time steps larger than 0.5 ns in Fig. \ref{2a}(a).

\subsection{Parameter dependency of skyrmion stretching}
The current density required to deform the skyrmion is smaller for small $J$ and $D$. Figure \ref{2a}(b) shows the change in maximum elongation for a given current when the LLG parameter $J$ and/or $D$ are either halved or doubled. For this simulation, the magnetic field is adjusted to the system to remain in the topological phase. Figure \ref{2a}(b) shows that for a constant ratio $J/D$, it keeps the same rate of elongation, while maintaining a similar shape as well. An alternative that allows for lower currents, is increasing $J$ and/or decreasing $D$. The skyrmion elongates more rapidly, but the downside is that it also grows wider and becomes more globular shaped.

Most researches try to maximize both DMI and exchange interaction to have robust devices against noise \cite{exchange1, exchange2, exchange3}. However in this work less might actually be better, because smaller $J$ and $D$ allow smaller currents to obtain skyrmion elongation. An advantage of skyrmions in thin films is that several parameters can be adjusted experimentally, for example, the strength of the DMI can be reduced by increasing the film thickness \citep{STT3}. 
C. Eyrich \cite{cobalt} showed that increasing the thickness of thin-film cobalt alloys can reduce the exchange constant.

\begin{figure}[h]
\includegraphics[width=0.8\columnwidth]{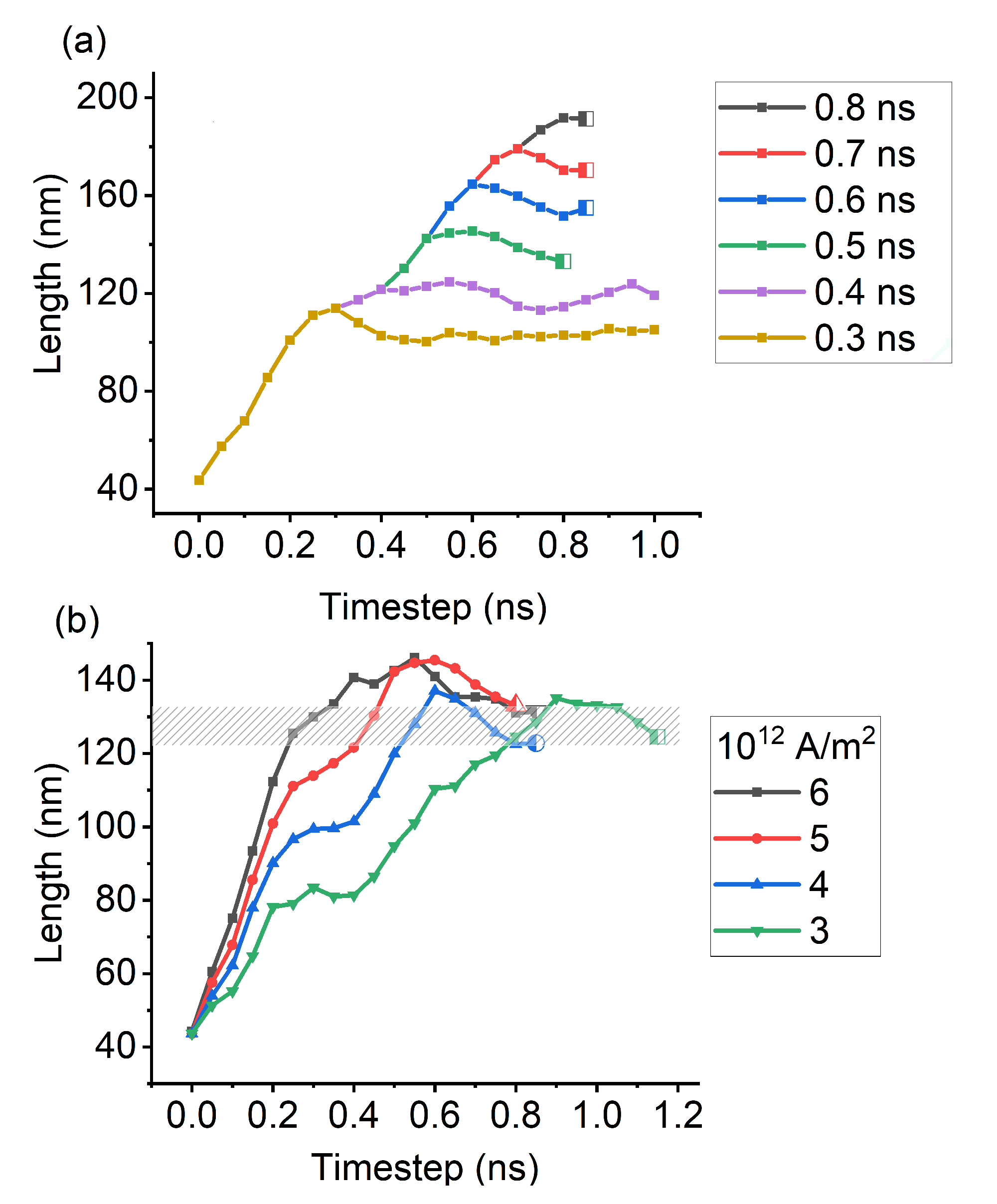}
\caption{(a) Skyrmion elongation evolution in time as function of different spin current pulse lengths. The applied current is fixed at $5 \cdot 10^{12}$ A/m$^2$. The half open dots indicate the skyrmion splitting point. If the skyrmion is elongated too much, it is unable to relax to its original configuration and splits in two. (b) Skyrmion elongation evolution in time as a function of current pulses strength, but differently from a) with current pulses selected for each case to be large enough to split the skyrmion. For each current the breaking point happens between 123-133 nm, indicated by the gray area. LLG parameters are $J = 2.5$ pJ/m, $D = 1$ mJ/m$^2$, $
\mu_0H = 0.1$ T  and $\alpha$ = 0.015.}
\label{2b}
\end{figure}

\subsection{Skyrmion splitting}

For long enough skyrmion deformation, a rupture of the skyrmion will follow after switching off the current. Figure \ref{2b}(a) illustrates the rate of elongation for increments in current pulses, but with a fixed current of $5\cdot10^{12}$ A/m$^2$. For short time pulses (up to 0.4 ns) the skyrmion will stretch out and relax back to its initial state over time. For larger time pulses, the length will reach a certain threshold after which it can no longer keep its elongated shape and ruptures in two as indicated by the half open dots. Figure \ref{2b}(b) shows the breaking points for different currents. The highest current (black curve) requires only a 0.4 ns current pulse, while the lowest current (green curve) needs a current pulse of 0.9 ns. In addition, they all show an average threshold length between 123 nm and 133 nm, as indicated by the gray area. This threshold length is dependent on the material parameters, as larger values of $J$ and $D$ will give a stronger binding, holding the skyrmion together. The remaining two skyrmions after the separation, will initially fluctuate in size, but they gradually relax to the size and shape of the original skyrmion.

\section{Stretched Skyrmion rotation}\label{sec:level4}

\begin{figure}[h]
\includegraphics[width=0.8\columnwidth]{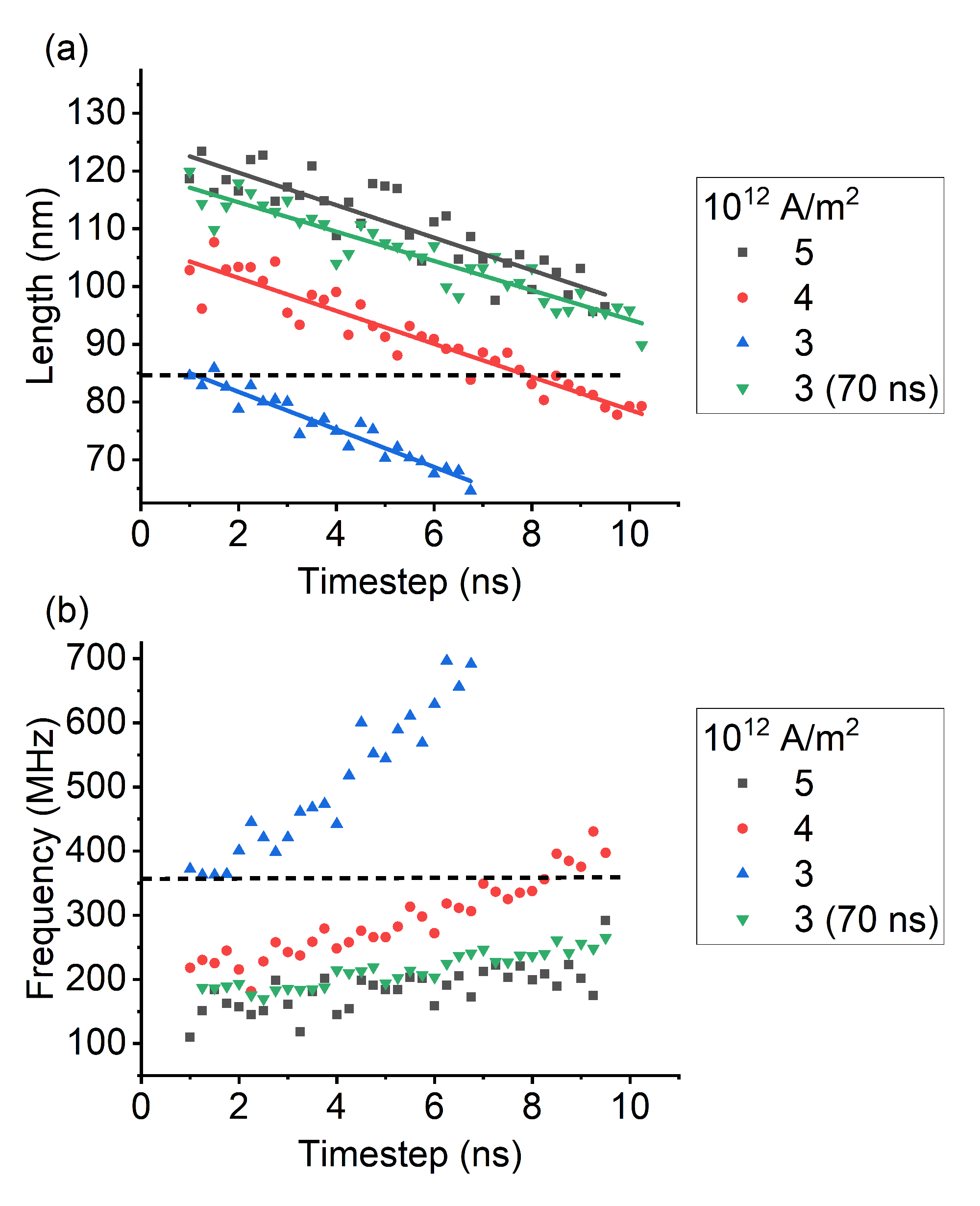}
\caption{(a) Skyrmion elongation evolution in time after different spin current pulses (0.4 ns current pulse for the black, red and blue curves and 0.7 ns for the green curve). (b) Frequency of the elongated skyrmion rotation versus the time for the same cases as in a). The dotted line shows where the length and frequency are equal. The LLG parameters are $J = 2.5\ pJ/m$, $D = 1 mJ/m^2$, $\alpha = 0.015$.}
\label{fig:long}
\end{figure}

\begin{figure}[h]
\includegraphics[width=0.8\columnwidth]{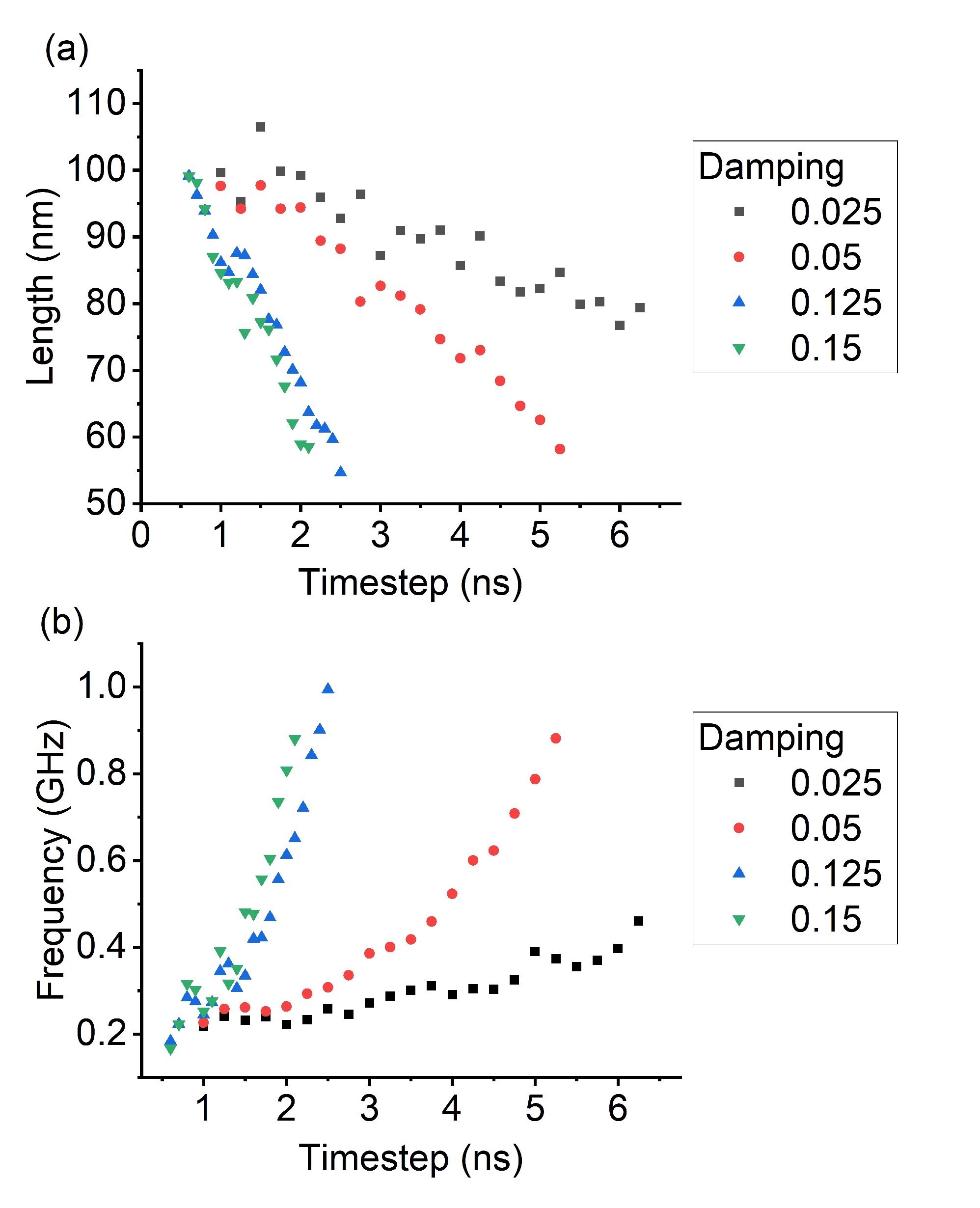}
\caption{(a) Skyrmion elongation relaxation from the initial conditions for different damping parameters on the LLG equation. Note how the skyrmion elongation decreases linearly but with a different slope depending on the damping. (b) Rotation frequency of the skyrmion in time for the same damping dependent relaxation processes as in a). The frequency increases in a parabolic manner while the radius decreases.
The LLG parameters are $J = 2.5\ pJ/m$, $D = 1 mJ/m^2$, $\alpha = 0.015$.}
\label{fig:long2}
\end{figure}

An important point is how the skyrmion behaves if it does not break. For short lengths, the elongated skyrmion rotates around its center with the direction of rotation determined by the external magnetic field orientation and the corresponding magnetization direction. Figure \ref{spin}(b) shows that with the used parameters, the rotation has a counter-clockwise motion. The length depends on the strength of the applied current and on the timed pulse duration, as shown in Fig. \ref{fig:long}(a). A current of $5\cdot10^{12}$ A/m$^2$ (black curve) with a time pulse of 0.4 ns has a similar length as a current of $3\cdot10^{12}$ A/m$^2$ (green curve) with a time pulse of 0.7 ns. Note too that the relaxation rate of the elongation is independent of the current strength or current pulse.

Figure \ref{fig:long}(b) shows that the frequency of rotation increases as the skyrmion elongation diminishes. We can see how the frequency is dependent on the elongation of the skyrmion for a given set of material parameters. In Fig. \ref{fig:long}(a) and \ref{fig:long}(b), we obtain for different strengths of the current, the same rotation frequency for the same elongation. As an example, some interesting values are pointed out in Fig. \ref{fig:long}(a) and \ref{fig:long}(b) with horizontal black dotted lines. The dotted line crosses the red curve at approximately $t=8$ ns and the blue curve at $t=1$ ns, for both the length and the frequency. This dependency of the frequency on the elongation hints at conservation of angular momentum.

Figure \ref{fig:long2}(a) shows the result of the simulations for a common set of parameters, but with different damping coefficients. The length shrinks faster with increasing damping coefficients, while the frequency increases. Secondly, the frequency, as illustrated in Fig. \ref{fig:long2}(b), no longer shows a linear trend, but is rather parabolic in shape.

\subsection{Momentum conservation}
The results from the previous section suggest that it has an analogy to the conservation of angular momentum for rotating skyrmions. For this conservation, the angular momentum at ${(t = 0)}$ should be equal to the angular momentum at any moment in time. For this to happen
\begin{equation} \label{eq:momentum}
    \left(\frac{r}{r_0}\right)^{-2} = \frac{\omega}{\omega_0}
\end{equation}
must be fulfilled with $r_0$ and $\omega_0$, corresponding to the initial elongation and frequency of the skyrmion.

According to Eq. \ref{eq:momentum}, when angular momentum is conserved, the relationship $(\omega/\omega_0)$ / $(r/r_0)^{-2} = 1$. In this case, the frequency of rotation increases as the length shrinks, while conserving energy. $(\omega/\omega_0)$ / $(r/r_0)^{-2} < 1$ would indicate a loss of angular momentum, while larger than one means a gain in angular momentum, with respect to conservation. 

\begin{figure}[ht]
\includegraphics[width=1\columnwidth]{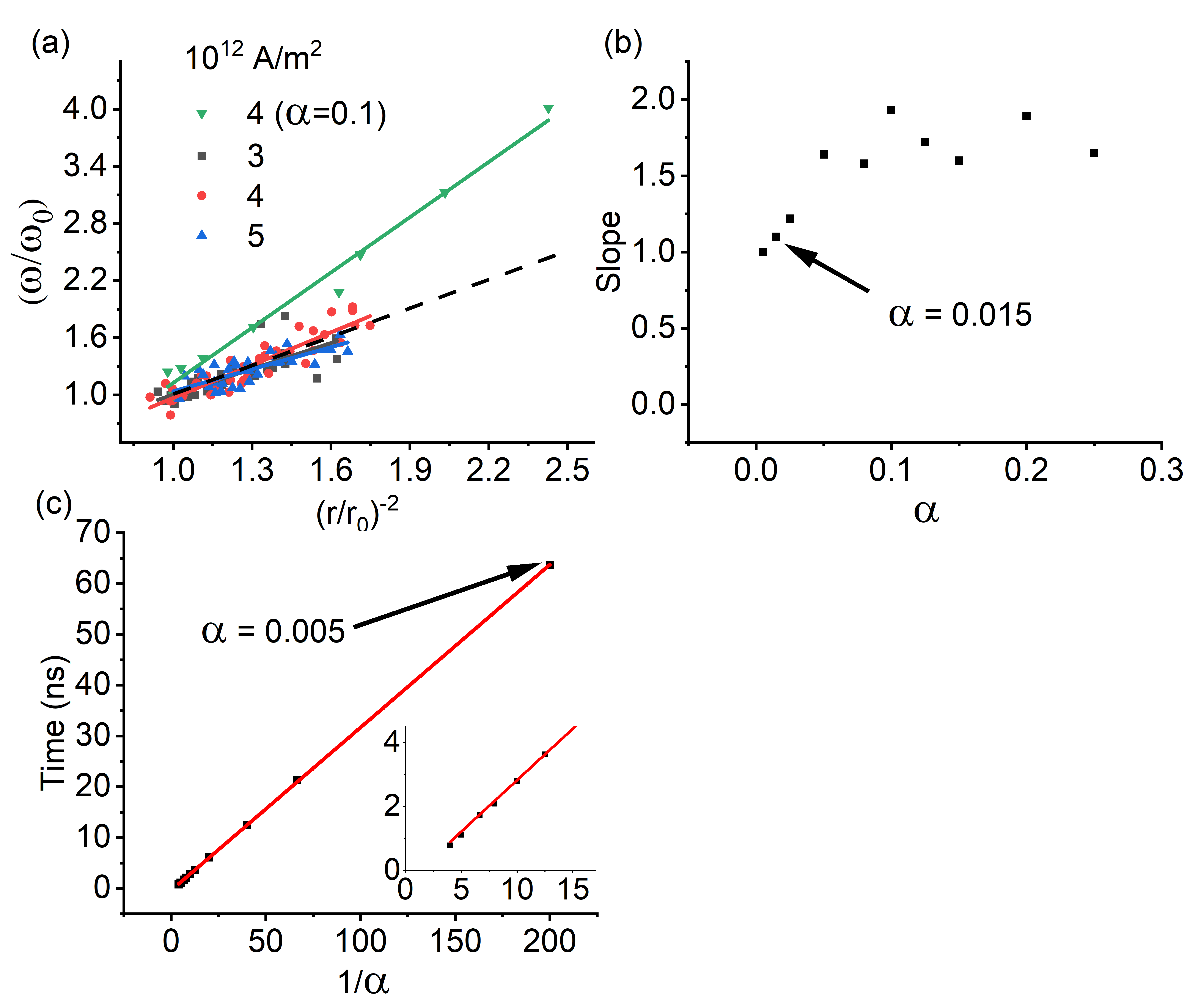}
\caption{(a) Shows  $(\omega/\omega_0)$ vs $(r/r_0)^{-2}$ and the different slopes for the currents 3, 4 and 5$\cdot10^{12}$ with a damping coefficient of 0.015 and one for $\alpha = 0.1$ (green curve). (b) The slope vs the damping coefficient is plotted for the current 4$\cdot10^{12}$. The slope saturates for higher $\alpha$ to just below 2. (c) Higher damping coefficient will damping out faster to their original size with a straight line according to 1/$\alpha$. }
\label{fig:damping}
\end{figure}

Figure \ref{fig:damping}(a) shows that for the physical damping coefficient of 0.015, the coefficient $(\omega/\omega_0)$ vs $(r/r_0)^{-2}$ for different currents is close to one, as shown by the black dashed line. This indicates that angular momentum is approximately conserved.

Additionally, a damping coefficient of 0.1 (green curve) has a slope of two. In Fig. \ref{fig:damping}(b) we see how the coefficient $(\omega/\omega_0)$ vs $(r/r_0)^{-2}$ is different from one for damping coefficients different than 0.015. The oscillatory behaviour, which can be seen at damping coefficients $\alpha>0.08$, is due to numerical errors, but it shows a slope between 1.5 and 2. For damping coefficients $\alpha > 0.25$, the system will relax too fast for any rotation to be observed. This gain in momentum is done a the expense of the total time it needs to return to the initial size. In Fig. \ref{fig:damping}(c) we can see how the relaxation time, that the elongated skyrmion needs to return to a circular shape, depends linearly on the inverse of the damping coefficient. The higher the damping, the larger the angular momenta attained in the relaxation process, but the relaxation time also becomes shorter.

The rotating behaviour of the skyrmion has the potential to be used as an oscillator. The speed of rotation can measured using a topological insulator as a heavy metal layer and measuring the change in resistance,  as described in \cite{measurement}.

\section{Conclusions}\label{sec:conclusion}
We have investigated the effect of applying a spatially varying spin current on the skyrmion. 
These two opposing current pulses stretch the skyrmion into an elongated shape, where the elongation length can be controlled by the strength of applied current and the pulse length. After the current pulse, the elongated skyrmion may either rotate or break, depending on its elongation length for a given set of parameters. 

The skyrmion splitting happens at a specific length threshold for different currents.
For relative short elongations, the skyrmion will rotate and its angular frequency can be controlled through the selection of specific elongations. The rate at which the skyrmion relaxes back to its circular shape depends on the damping coefficient.

To conclude, the proposed device uses two opposite spin currents to duplicate skyrmions by means of stretching it, or cause the elongated skyrmion to rotate. These phenomena could be used to create devices to generate more skyrmions and it can be used as a skyrmion oscillator in the MHz-GHz range.

\begin{acknowledgments}
We acknowledge the Horizon 2020 project SKYTOP “Skyrmion-Topological Insulator and Weyl Semimetal Technology” (FETPROACT2018-01, n. 824123). We also like to thank Sandra Vetter for her contribution in the graphical design.
\end{acknowledgments}


\bibliography{apssamp}

\end{document}